\documentclass[12pt,preprint]{aastex}

\slugcomment{\today}

\shorttitle{Coronal Abundance of Iron}
\shortauthors{White, Thomas, Brosius, Kundu}

\begin{document}

\title{The Absolute Abundance of Iron in the Solar Corona}

%% Use \author, \affil, and the \and command to format
%% author and affiliation information.
%% Note that \email has replaced the old \authoremail command
%% from AASTeX v4.0. You can use \email to mark an email address
%% anywhere in the paper, not just in the front matter.
%% As in the title, you can use \\ to force line breaks.

\author{S. M. White\altaffilmark{1}, R. J. Thomas\altaffilmark{2}, J. W.
Brosius\altaffilmark{2,3} \& M. R. Kundu\altaffilmark{1}}
\altaffiltext{1}{Astronomy Department, University of Maryland, College Park, MD 20742}
\altaffiltext{2}{NASA/Goddard Space Flight Center, Code 682, Greenbelt MD 20771}
\altaffiltext{3}{Raytheon ITSS, 4400 Forbes Boulevard, Lanham, MD 20706}
%\email{white,jerry,kundu@astro.umd.edu}

\begin{abstract}
We present a measurement of the abundance of Fe relative to H in the solar
corona using a technique 
which differs from previous spectroscopic and solar wind measurements.
Our method combines EUV line data from the CDS spectrometer
on {\it SOHO} with thermal bremsstrahlung radio data from the VLA. The
coronal Fe abundance is derived by equating the thermal
bremsstrahlung radio emission calculated from the EUV Fe line data to
that observed with the VLA, treating the Fe/H abundance as the sole
unknown.
We apply this technique to a compact cool active region and find Fe/H
 = 1.56 $\times$ 10$^{-4}$, or about 4 times its value in the solar
photosphere. Uncertainties in the CDS radiometric calibration,
the VLA intensity measurements, the atomic parameters, and the
assumptions made in the spectral analysis yield net uncertainties
$\sim$ 20\%.  This result implies that 
low first ionization potential elements such as Fe are enhanced in the solar
corona relative to photospheric values. 
%A larger Fe abundance also 
%helps to explain difficulties found in 
%earlier comparisons between radio and EUV/soft X--ray data.

\end{abstract}


\section{INTRODUCTION}

It seems remarkable that the absolute abundances (here taken to mean
abundances relative to hydrogen) of trace elements in the
solar corona are still a matter for debate when many 
measurements for much more distant stars exist. The reason for this situation
is easy to understand. Classic spectroscopic techniques for measuring
relative abundances require that we compare the amount of radiation
emitted in spectral lines whose
atomic physics we understand and which arise from the two different elements
in question.
In the solar corona H and He are fully ionized, and hence produce no lines. In
stars with hotter coronae, thermal bremsstrahlung produces a continuum
emission which can be used instead of a spectral line as a measure of the
amount of H present, but the Sun's quiescent corona is so cool that the
continuum present at X--ray or EUV wavelengths is relatively weak and
difficult to measure.

It has been known for some time that there is a difference in the
relative abundances of elements in the photosphere and the corona organized
according to the first ionization potential (FIP) of the element:
high--FIP elements such as O are underabundant in the corona relative to
low--FIP elements such as Fe by a factor of 4 \citep{Mey85}. However, it has
not been clear whether low--FIP
elements are enhanced in the corona in an absolute sense (i.e., as a
fraction of the total ion number density, dominated by H  and He) or the
high--FIP elements are depleted relative to H. Two main techniques have been
used to infer absolute abundances in the solar
corona \citep[]{FSH99}. One can observe solar flares, which produce plasma hot
enough for the continuum to be measured, and compare the continuum with a
suitable set of lines; or one can measure particles directly in
the solar wind, and assume that the abundances there truly reflect the
abundances
in the solar corona. Unfortunately, these two techniques have resulted in
conflicting measurements for, e.g., the absolute abundance of iron. The flare
measurements suggest that it is close to the absolute abundance in the
photosphere \citep{VeP81} or perhaps factors of 1.5 -- 2 
larger \citep{FlS99}, while the
solar wind particle measurements suggest that it is 4 times the photospheric 
ratio \citep{Rea92}. 

In this paper we use a third technique to measure the absolute abundance
$N_{\rm Fe}/N_{\rm H}$ which has several desirable features. We use EUV 
observations of lines
of Fe in order to determine the amount of Fe present in the corona, but
to measure the amount of H present we use radio continuum measurements of
thermal bremsstrahlung from coronal plasma. This measurement therefore pertains
to the quiescent solar corona, including non-flaring weak active regions,
rather than flare or solar wind plasma. Like
classic spectroscopic techniques, this technique should be independent of the
filling factor of plasma in the volume studied since both types of emission
have the same dependence on number density. We describe the measurement
technique and the observations used for this measurement in the following section.
We then discuss the implications of our measurement.

\section{THE MEASUREMENT TECHNIQUE}

The method proposed here to measure the abundances of elements relative
to H rests on the fact that the EUV line emission and
radio bremsstrahlung both depend on the product of the
electron number density and the number density of an element. In the case of
radio
bremsstrahlung the element is effectively hydrogen; in the case of an
EUV line, it is the element responsible for the line. Therefore
comparison of the
two yields the abundance of the latter element relative to hydrogen.

Specifically, since
most EUV lines produced in the solar corona are collisionally excited
and optically thin, the
line intensity at a photon energy $E$ corresponding to a transition in
ionization state
Z of element X is the integral over temperature of the product 
$Q_{X,Z,E}\,DEM(T)\,{N_X/N_{\rm H}}$ \citep{FSH99}, 
where $Q_{X,Z,E}$ is the ``contribution function'': it includes the
intrinsic spontaneous emission rate (the Einstein ``A'' coefficient)
as well as the
ratio $N_{X,Z}/N_X$ of the number density of the particular charge state
$Z$
of the element $X$ responsible for the line to the total
number density of the element. This ratio is usually determined by an
ionization balance calculation appropriate to
the temperature of the plasma, and $Q$ is a strong function of
temperature
for this reason. However, for most lines $Q$ is effectively independent
of $N_X$ and $N_e$. The (hydrogen column) 
differential emission measure DEM(T) is
the integral along the line of sight of ${d(N_e\,N_{\rm H})/dT}$.

Bremsstrahlung is one of the two main mechanisms for radio
emission in the solar corona
and also has a significant dependence on
temperature: in the optically thin limit, the flux at a frequency $f$ is

\begin{equation}
 S \ = 9.78\,\times\,10^{-3}\,{2\,k_B \over c^2} (1 + 4 {N_{\rm He} \over N_{\rm
H}}) \ \int \int \ T^{-0.5}\,DEM(T)\,G(T) \ dT \,d\Omega 
\end{equation}

\noindent in ergs cm$^{-2}$ s$^{-1}$ Hz$^{-1}$ \citep{Dul85},
where $G(T)$ = 24.5 - ln(${T/f}$) is the Gaunt factor, $k_B$ is the
Boltzmann constant, $c$ the speed of light and the second integral is over the
solid angle subtended by the source.
We are assuming that in a plasma at the temperature of the solar corona the ion 
component is
completely dominated by protons and fully--ionized helium, and we adopt
$N_{\rm He}/N_{\rm H}\,=\,0.1$. It is clear from (1) that we must know the 
DEM in order to calculate the thermal bremsstrahlung radio flux accurately.
Note that as long as $N_X/N_{\rm H}$ does not vary from
place to place, (1) has exactly the same dependence on density as the
EUV line emissivity and
thus filling factor has exactly the same effect on the EUV line and radio 
continuum fluxes.

We determine the abundance $N_{\rm Fe}/N_{\rm H}$ with the following steps: 
(i) We measure the radio flux from a source in the solar
corona. This source {\it must be optically thin} and its radio flux 
{\it must be due only to bremsstrahlung}. 
The latter requirement can be tested by
observing the radio spectrum of the source, which should be essentially flat
for optically thin bremsstrahlung. (ii) We observe the same region with the
Coronal Diagnostic Spectrometer (CDS) on {\it SOHO} in order to measure the
intensities of lines from a range of charge states of Fe
sufficient to give a reliable determination of the DEM. 
%CDS has a powerful
%combination of spectral and spatial resolution which makes this study possible.
(iii) We use the CHIANTI atomic data package \citep{DLM97,LLD99} to fit
the Fe line fluxes to a DEM.
(iv) We assume that the shape of the DEM for electrons and protons is the same
as that for Fe, and integrate over temperature in (1) to 
predict the optically--thin
bremsstrahlung radio flux we expect to observe, for a given absolute abundance
of Fe. Comparison with the radio flux measured by the 
VLA yields $N_{\rm Fe}/N_{\rm H}$.

Coordinated observations with CDS and the VLA\footnote{The Very Large Array is
run by the National Radio Astronomy
Observatory which is a facility of the National Science Foundation operated
under cooperative agreement by Associated Universities, Inc.} 
were carried out on 1997 July
27, Aug.~3 and Nov.~11. We will present an analysis of all the data elsewhere;
here we discuss the best target observed, a small active region AR 8105
found at about S35W15 on 1997 Nov.~11 (with the VLA in ``D''
configuration). This was the best target in the sense that it was
compact: the VLA is an interferometer and cannot measure the flux of large
sources, so radio flux measurements are more reliable for compact sources. The
region was one of four  observed on this day 
with the VLA at 1.4, 4.8 and 8.4 GHz. CDS also
observed the same 4 regions alternately using the sequence FE\_IONS which
reads nine 21--pixel--wide windows from the NIS1 detector of CDS containing 
14 identified lines including those of Fe X, XI, XII, XIII, XIV, XV, XVI and XVII.
One run of the sequence took 32 minutes, scanning with the 4\arcsec\ slit
across a 240\arcsec\ wide field of view. Five sequences were acquired on AR
8105 centered at about 13:14, 16:28, 18:42, 20:56 and 23:10 UT, respectively. 
The VLA
observations covered the period from 14:30 UT to 23:10 UT, cycling between the
different frequencies rapidly in order to achieve optimal $uv$ coverage.
%At 4.8 and 8.4
%GHz where the VLA field of view does not cover the whole solar disk the
%observations were timed to coincide with the region being observed concurrently 
%by CDS. 
Images of AR 8105 acquired by CDS, the VLA and EIT are shown in Figure 1.

\section{THE ABUNDANCE OF IRON}

It is essential for this measurement that the VLA see the same plasma
that is seen in the EUV, and Fig.~1 shows that this is indeed the case: the
EUV and radio images match each other extremely well, indicating that
gyroresonance opacity does not contribute significantly to the radio emission
and we conclude that it is due to bremsstrahlung. The higher--resolution
EIT images shows that two east--west loops dominated AR 8105.

The VLA observations resulted in images with resolutions of 30\arcsec\
at 1.4 GHz, 13\arcsec\ at 4.8 GHz and 7\arcsec\ at 8.4 GHz.
Using the 4.8 and 8.4 GHz images we identified an area of dimension 
69\arcsec\ $\times$ 97\arcsec\ around AR 8105 which includes essentially all
the flux from the region.  Radio fluxes were derived by summing over this area at
4.8 and 8.4 GHz, while due to the poorer resolution at 1.4 GHz we instead
fitted a Gaussian model to the feature associated with AR 8105 and determined
its flux. At each frequency we have independent measurements in
two separate 50 MHz bands and in the two circular polarizations, all of which
are consistent with one another. The resulting fluxes are 0.25 $\pm$ 0.01 sfu
at 1.4 GHz, 0.207 $\pm$ 0.004 sfu at 4.8 GHz and 0.197 $\pm$ 0.006 sfu at 8.4
GHz, which are consistent with the $f^{-0.1}$ spectrum expected for
optically thin bremsstrahlung due to the Gaunt factor in (1). 
The quoted uncertainties are the map rms per
beam times the square root of the number of beams over which integration was
carried out, and are much larger than the formal uncertainties in the fluxes.
Polarizations were also low, consistent with the properties of bremsstrahlung.

The CDS data were calibrated using the standard reduction package in SOLARSOFT
(results presented here use the version current in 1999 December). Line profiles
were fitted and fluxes extracted, with uncertainties calculated according to
\citet{Tho98}. We used both a recent radiometric CDS calibration derived from 
comparison with the SERTS rocket flight in 1997 \citep{TTK00} 
and the ``version 2''
CDS calibration which was in the CDS software as of 1999 December.
One important step is that background subtraction was carried
out on the CDS data using a linear fit to pixels to the north and south 
of the region studied, well away from AR 8105. This step is important because of the
fact that the VLA does not measure large--scale flux and hence is insensitive
to the quiet--Sun contribution. In effect we assume that a smooth quiet--Sun
contribution has been subtracted from the radio fluxes, and
hence it should be subtracted from the CDS data as well. This predominantly
affects the cooler lines, amounting to up to 20\% of the photons in the
original spectrum, but being fairly uniform across the CDS field of view.

The following lines (with typical line fluxes averaged over the region 
shown in parentheses, in ergs cm$^{-2}$ s$^{-1}$ st$^{-1}$) 
were then used to calculate
the DEM: Fe X at 345.7 \AA\ (55 $\pm$ 3), Fe XI at 352.7 \AA\ (140 $\pm$ 4), 
Fe XII at 352.1 \AA\ (145 $\pm$ 4) and 364.4 \AA\ (230 $\pm$ 4),
Fe XIII at 348.3 \AA\ (95 $\pm$ 4), Fe XIV at 334.2 \AA\ (260 $\pm$ 4), 
Fe XVI at 360.7 \AA\ (380 $\pm$ 4) and Fe XVII
(undetected here) at 350.5 \AA. These lines were chosen to be as
density--insensitive as possible while still
covering the temperature range expected for the non--flaring solar corona,
log T = 5.8 up to 6.8. The Fe XIII and Fe XII lines do have significant
sensitivity to density in the likely range of coronal densities, 10$^9$ --
10$^{10}$ cm$^{-3}$. In practice, AR 8105 was quiet and there
is no significant plasma outside the temperature range 5.9 -- 6.4. The fitting
procedure uses bins of width 0.1 in log T; in effect, we find that only 4 bins
contain significant emission measure and the adjacent two bins are also
significant, so we are fitting to 6 parameters using 8 lines from 7 different 
charge states. The Fe XV line at 327.1 \AA\ was not used to
determine the DEM because it 
was always found to be $\sim$ 30\% lower than the models
predicted; the atomic data for it are clearly not
consistent with the strong Fe XIV and XVI lines observed. 
Additional details will be given in a companion paper.

We adopted the ionization equilibrium calculations of \citet{ArR92} and assume
a constant pressure $nT$ of 1 $\times$ 10$^{16}$ cm$^{-3}$ K (e.g., $N$ = 5
$\times$ 10$^9$ cm$^{-3}$ at $T$ = 2 $\times$ 10$^6$ K) in
determining the DEM.
The line fluxes used for the fit were generally reproduced by the
resulting model DEM to better than 10\% (SERTS) or 5\% (``version 2'').
Figure 2 shows the geometric mean DEM over the 5 CDS sequences
with error bars showing the standard deviation in each temperature bin.
Essentially no plasma is present below log T = 6.0 (controlled by the ratio of
Fe X to Fe XI) while the amount of plasma above 2 $\times$ 10$^6$ K is small
(controlled by the ratio of Fe XVI to Fe XIV).
These properties are consistent with those of DEMs derived for other regions
of similarly low activity \citep{BDT96}.
Integrating over each of the 5 DEMs resulted in a predicted radio flux of 0.829
$\pm$ 0.023 sfu in the 6700 arcsec$^{2}$ area at 4.8 GHz if we use the
CDS calibration based on the SERTS--97 comparison and adopt
$N_{\rm Fe}/N_{\rm H}$ = 3.9 $\times$ 10$^{-5}$ as in \citet{Mey85}, 
which is close to the photospheric value. 
The observed radio flux of 0.207 sfu 
is a factor of 4.0 smaller than this. Consequently we derive a value of 1.56
$\times$ 10$^{-4}$ for $N_{\rm Fe}/N_{\rm H}$, or $\log_{10}{A_{\rm Fe}}$ = 8.19 on 
the commonly used scale where $\log_{10}{A_{\rm H}}$ = 12.00. 
Using the ``version 2'' CDS calibration we predict 0.716
$\pm$ 0.013 sfu and $N_{\rm Fe}/N_{\rm H}$ = 1.35 $\times$ 10$^{-4}$
($\log_{10}{A_{\rm Fe}}$ = 8.13). 
We adopt the more recent SERTS--97 calibration for our final result.
The uncertainty of 3\% for the predicted radio flux is the standard error in
the mean of the results of the five distinct CDS observations of AR 8105 and
thus incorporates the effects of both temporal variability over the 12 hours
of the experiment as well as the stochastic nature of the process of fitting
the line fluxes to a DEM. We adopt an uncertainty of 15\% for the
CDS radiometric calibration \citep{TTK00}.

It turns out that the measured value for $N_{\rm Fe}/N_{\rm H}$
is reasonably robust to several of the
assumptions made here, mostly because it depends on an integration over the
DEM and is thus not very sensitive to details of the DEM. 
One assumption is that the result quoted uses the version of
CHIANTI (0.9) used by the CDS team at NASA/GSFC as of 1999 December, which is not
the most current version (2.0). For the Fe lines used here the only
differences between the two versions appear to be in Fe X and Fe XI, and the
newer version results in significantly poorer fits for the DEM because of these
two lines: the resulting DEMs have
less emission measure at low temperatures and more at high temperatures.
Notwithstanding this fact, the result for the abundance differs by less than
3\% with the newer version of CHIANTI. If we use the ionization equilibrium
calculations of \citet{ArR85}, the DEMs show a pronounced peak at $\log{T}$ =
6.3 and the resulting abundance is about 10\% larger.
A change in the assumed pressure by a factor of 2 affects the results 
by a similar amount: it would be 8\% smaller at 5.0 $\times$
10$^{15}$ cm$^{-3}$ K and 10\% larger at 2.0 $\times$ 10$^{16}$
cm$^{-3}$ K. We attempt to minimize the uncertainty due to variation in
pressure across the active region by
choosing a value representative of the source region for the
bulk of the EUV line emission.
We note also that the background subtraction carried out
on the CDS data introduces some uncertainty in the final results, as 
does solar calibration at the VLA, which uses dedicated noise sources for
solar calibration which need to be measured routinely to maintain calibration
accuracy. However, the calibration at each of the three frequencies observed is 
independent of the others, so we take the fact that the measured spectrum is
flat to be an indication that solar calibration was not in error by more than
the 20\% levels which all the other factors may introduce.
%
%Another possible source of uncertainty is the fact that we have averaged the
%CDS spectrum over the whole region before fitting to a DEM. We use a region
%larger than AR 8105 itself because of the problem that the images
%have different resolutions and thus flux comparisons become difficult; 
%the region then necessarily includes areas where the background--subtracted
%EUV line fluxes are small and large errors are likely to result in fitting a DEM.
%We have checked using smaller bright 
%regions that fitting the DEM on a pixel--by--pixel
%basis yields the same final answer for the predicted radio flux as fitting the
%line fluxes averaged over a number of pixels, to within a few percent.

\section{CONCLUSIONS}

Our measurement of an absolute abundance for Fe which is 4 times
photospheric is clearly consistent with the interpretation that low--FIP
elements such as Fe are enhanced relative to their photospheric abundances. Of
the commonly--used solar abundance tables, Feldman's (1992) \nocite{Fel92} is
closest to our measurement with $\log_{10}{A_{\rm Fe}}$ = 8.10; his value  is
based more on the argument that H should be treated as a high--FIP element rather
than any specific measurement of $N_{\rm Fe}/N_{\rm H}$. 
Other commonly used values for $\log_{10}{A_{\rm Fe}}$ are the photospheric
results 7.60 \citep{All73}, 7.51 \citep{AnG89} and 7.67 \citep{GrA91}, the
coronal results 
7.59 \citep{Mey85}, 7.83 \citep{FlS99} and 8.50 \citep{WMD94}, and the
solar wind measurements 7.93 \citep{Rea95} and 8.51 \citep{Rea99}.
%
%An important consequence of this measurement is that emission measures, and
%hence measures of energy, derived using wavelength bands dominated by coronal
%lines of Fe are too large by a factor of order 4.0 if an abundance near 7.60
%is used. In particular this has been true of nearly all quantitative analysis of {\it
%Yohkoh} SXT spectra (see discussion by Schmelz et al.~1999)
%since the analysis package assumes Meyer coronal abundances \nocite{SSS99}. 
%We also note that a larger coronal Fe abundance resolves a
%long--standing puzzle which has arisen in comparisons of solar radio data with
%EUV and soft X--ray measurements: using emission measures and temperatures
%derived from the latter measurements, virtually all such studies prior to 1993
%found that the bremsstrahlung radio flux expected from just the hot
%component of the corona over active regions was more than the radio
%telescopes observed, by a factor of several (see discussion by White 2000).
%\nocite{Whi00} This result led to the interpretation that there must be unseen
%cold absorbing plasma present which acted to diminish the radio flux.
%However, all these studies used Fe abundances close to photospheric, and the
%difficulties ensuing are removed if a larger Fe abundance is appropriate
%(e.g., Brosius et al. 1993). \nocite{BDT93}
%Comparison of EIT data with radio images is also consistent with the larger
%value for the Fe abundance \citep{ZWK99}. 
We note in passing that
spectroscopic measurements of the Fe abundance in active stars with coronae
much hotter than the Sun's have tended to suggest abundances smaller than
photospheric \citep{MKW96,Dra96}, in contrast to the solar behaviour we find
here.

We have reported here a single measurement of the Fe abundance. Spectroscopic
studies of relative abundances suggest that they may vary considerably from
flare to flare and from one region to the next \citep{SLM84,Sab95,SLB98}, so
it is clearly important to confirm our measurement. We have additional data
sets which will be published in a longer paper elsewhere, but with larger
uncertainties expected in the radio fluxes, and expect to carry out further
studies of this type in the future.

\section*{Acknowledgments}

This research at the University of Maryland was carried out primarily under
a SOHO GI grant from NASA, NAG 5--4954. Solar radiophysics at the University
of Maryland is also supported by
NSF grant ATM 96-12738 and NASA grants NAG 5-7370, NAG
5-6257 and NAG 5-7901. We thank Richard Harrison, Andrzej Fludra and
the CDS operators for their assistance with the project, and Barry Clark of
NRAO for scheduling the VLA observations to match those of SOHO.
We thank the CHIANTI consortium for their efforts to make atomic data
readily accessible.

\clearpage

%%
%\bibliographystyle{/home/white/biblio/natbib/apj}
%%
%\bibliography{/home/white/biblio/solar,/home/white/biblio/stellar}
% 
%\typeout{ }
%\typeout{!!!!!!!!!!!!!!!!!!!!!!!!!!!!!!!!!!!!!!!!!!!!!!!!!!!!!!!!!!!!!!!}
%\typeout{        BIBDATA NEED TO BE READ IN}
%\typeout{!!!!!!!!!!!!!!!!!!!!!!!!!!!!!!!!!!!!!!!!!!!!!!!!!!!!!!!!!!!!!!!}
%\typeout{ }

\clearpage

\begin{figure}
\figurenum{1}
\plotone{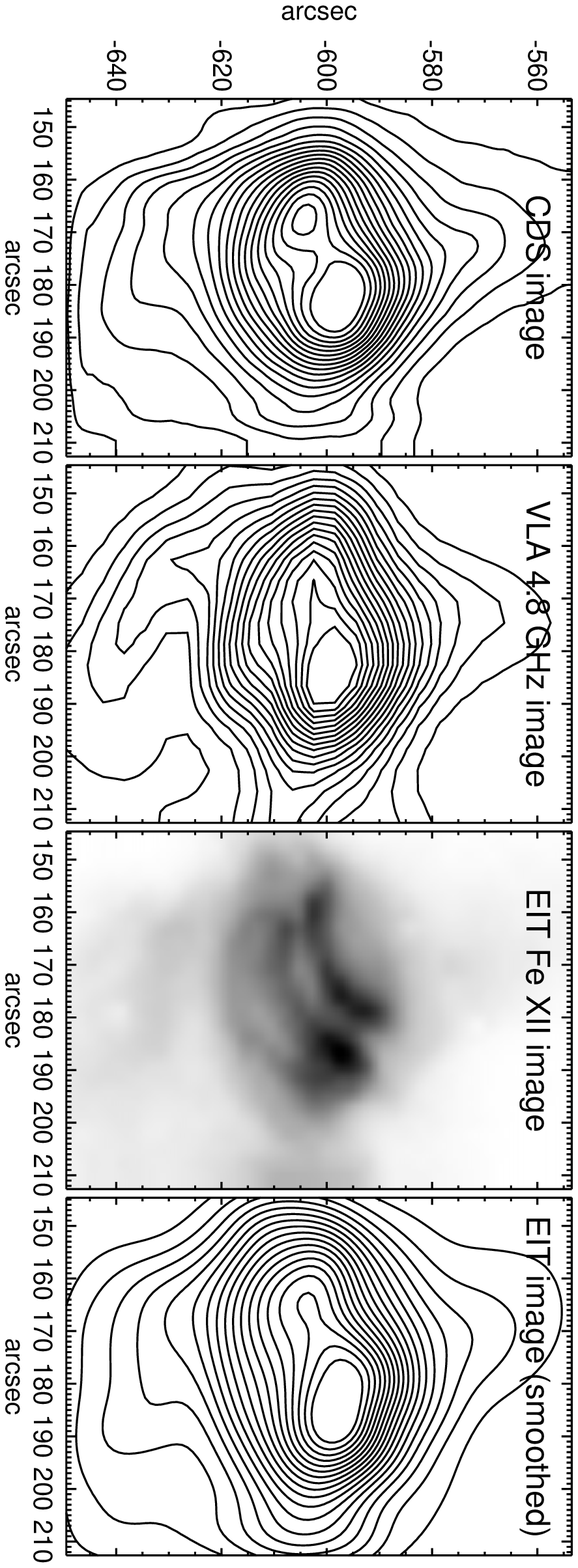}

\figcaption{Images of the target region used for this study obtained by CDS
(left panel, all photons, smoothed with a 7\arcsec\ gaussian), VLA (second
panel, 4.8 GHz, beam size 13\arcsec), EIT Fe XII 195 \AA\ (third panel,
greyscale, 5\farcs2 resolution) and the EIT image smoothed to 10\arcsec\
resolution and contoured for direct comparison with the CDS and VLA images.
The CDS and EIT images have been background--subtracted as described in the
text. In the contour images, contours are plotted at 5\%, 10\%, ..., 95\% of the
maximum for that image.
}
\end{figure}

\clearpage

\begin{figure}
\figurenum{2}
\epsscale{0.90}
\plotone{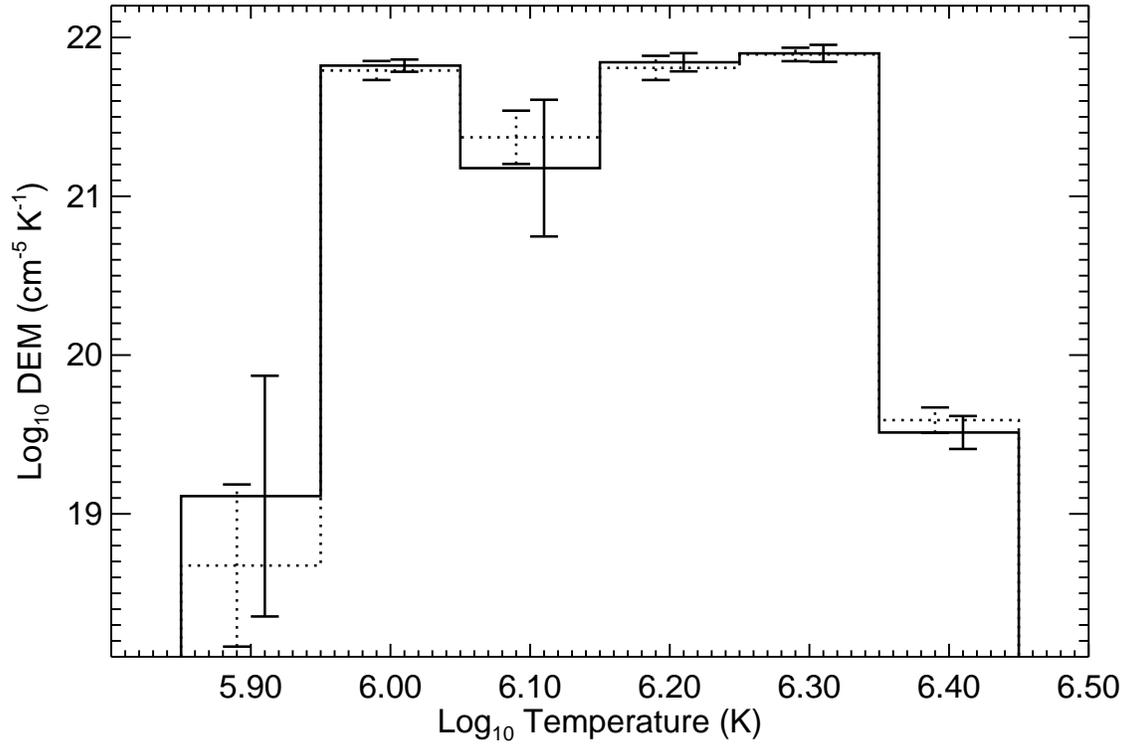}

\figcaption{
The geometric mean of the best--fit differential emission measure
distributions for the five CDS observations of the region studied.
The solid line shows the results for the SERTS 1997 calibration of CDS,
while the dotted line shows the results for the ``version 2'' calibration.
Error bars show the standard deviation (in the log) 
over the five measurements in each temperature bin.
\label{fig2}} 

\end{figure}

\end{document}